\documentclass[a4paper, notoc, 12pt]{article}
\pdfoutput=1
\usepackage{jcappub}
\usepackage{feynmp}
\usepackage{graphicx}  
\usepackage{bm}        
\usepackage{amssymb}   
\usepackage{slashed} 
\usepackage{amsfonts}
\usepackage{wasysym}
\usepackage{amssymb}
\usepackage{hyperref}
\usepackage{textcomp}
\usepackage[caption=false]{subfig}
\usepackage{braket}
\usepackage{placeins}

 \newlength{\wth}
\newcommand{\be}{\begin{equation}}
\newcommand{\ee}{\end{equation}}
\newcommand{\bea}{\begin{eqnarray}}
\newcommand{\eea}{\end{eqnarray}}

\newcommand{\beqa}{\begin{eqnarray}}
\newcommand{\eeqa}{\end{eqnarray}}

\newcommand{\beq}{\begin{equation}}
\newcommand{\eeq}{\end{equation}}

\newcommand{\dn}{\Delta N_{\rm eff}}

\newcommand{\La}[1]{{\ensuremath{\Lambda_{\rm #1}}}}

\title{
The Darkness of  Spin-0 Dark Radiation }
\author[]{M.C.~David Marsh}
\affiliation[]{Rudolf Peierls Centre for Theoretical Physics, University of Oxford,\\ 1 Keble Road, Oxford OX1 3NP, United Kingdom}
\emailAdd{david.marsh1@physics.ox.ac.uk}

\abstract{
 We show that the
scattering of a general spin-0 sector of dark radiation off the pre-recombination thermal plasma results in 
  undetectably small spectral distortions of the Cosmic Microwave Background.  
 
}

\begin{document}


\maketitle

\section{\label{sec:intro} Introduction and Conclusions}

The existence of dark radiation, i.e.~an additional contribution to the relativistic energy density of the universe beyond that of the photons and the neutrinos of the Standard Model, 
\be
\rho_{\rm d.r.} = \rho_{\rm rad.}^{\rm tot} - \rho_{\gamma} - \rho_{\nu} \, ,
\ee
is a a topic of considerable recent
 observational and theoretical interest. 
Conventionally, $\rho_{\rm d.r.}$
is parametrized 
in excess units of sterile neutrino species, $\dn$, through the equation
\be
\dn = \frac{8}{7} \left( \frac{11}{4}\right)^{4/3} \frac{\rho_{\rm d.r.}}{\rho_{\gamma}} \, .
\ee 
%
%
 As dark radiation contributes to the total energy density of the universe, it 
leads to an increase in the Hubble expansion rate. The 
primordial helium mass fraction, $Y_p$, is quite sensitive to the expansion rate during BBN and may --- in combination with the primordial deuterium abundance --- 
be compared with the observationally inferred primordial light element abundances to provide constraints on the baryon-to-photon ration $\eta$ and $\dn$. 
By this method, recent BBN determinations of $\dn$  give \cite{cooke_precision_2013},
\be
\dn \big|_{\rm BBN} = 0.46\pm0.20 \, ,
\ee
at the 68\% confidence level. Similarly, using the data from \cite{izotov_primordial_2013, pettini_new_2012}, reference \cite{nollett_bbn_2014} found 
\be
\dn \big|_{\rm  BBN} =0.51\pm0.23 \, .
\ee

Independently, 
the amount of dark radiation present during  the epoch of  recombination can be constrained by 
the precise distribution of the 
CMB anisotropies. 
An increased expansion rate 
during recombination results (upon keeping the angular size of the sound horizon, $\theta_s$, fixed) in an increased Silk damping of the high-$l$ multipoles \cite{hou_how_2013}.  The best-fit estimates 
provided by the Planck collaboration are,
\be
\dn \big|_{\rm  CMB} =0.26\pm0.27 \, ,
\ee
for  (Planck+WP+highL+BAO) and, 
\be
\dn \big|_{\rm  CMB} =0.48\pm0.25 \, ,
\ee
for the combination (Planck+WP+highL+BAO+H$_0$) that includes the measurement of the locally observed value of the Hubble constant \cite{planck_collaboration_planck_2013}.\footnote{In \cite{planck_collaboration_planck_2013}, a $2.5\sigma$  discrepancy between the $\Lambda$CDM best-fit value of $H_0$ and the locally inferred value of \cite{riess_3_2011} was found. Since the announcement of the results of the Planck experiment, an updated version of \cite{planck_collaboration_planck_2013}
 accounts for a previously unnoticed systematic error and a consequent increased best-fit $\Lambda$CDM value of $H_0$ by $0.3\sigma$. Furthermore, by including a proposed distance recalibration of \cite{humphreys_toward_2013}, an updated version of \cite{riess_3_2011} has been found to give an $0.33\sigma$ decrease in the locally observed value of the Hubble constant \cite{riess_adam_g._local_2014}. The consistency of the CMB  fitted value of $H_0$ within the $\Lambda$CDM model, baryon acoustic oscillations and the observations of the local universe are discussed in \cite{bennett_1_2014}.} \footnote{If a tensor-to-scalar ratio of $r\approx 0.2$ is assumed \cite{bicep2_collaboration_bicep2_2014}, then the inferred best-fit-values from the CMB analysis is significantly modified to  $\dn \big|_{CMB} = 0.81\pm0.25$ \cite{dvorkin_neutrinos_2014, giusarma_relic_2014}.}

In addition, upon assuming that  $\dn$ remains constant between BBN and recombination,  
a joint analysis of the  
CMB data from the Planck collaboration and the inferred primordial deuterium abundance gives \cite{cooke_precision_2013},
\be
\dn \big|_{\rm  BBN+ CMB} =0.23\pm0.28 \, .
\ee
By including the recent observational determination of $Y_p$ \cite{izotov_primordial_2013}, 
reference  \cite{nollett_bbn_2014} found,
\be
\dn \big|_{ \rm BBN+ CMB} = 0.40 \pm0.17 \, .
\ee
In sum, both the CMB and BBN analysis give inconclusive hints of the existence of dark radiation at the $1$--$2.5\sigma$ level.  The Planck experiment is expected to continue to improve the bounds on $\dn$ with a projected sensitivity of $\pm 0.2$ \cite{hamann_using_2008}. Future CMB polarisation experiments may obtain a sensitivity to $\dn$ of $\pm 0.044$ \cite{galli_constraining_2010}.

 A conclusive  observation of $\dn > 0$ 
  could be explained by  
    a higher energy density in the neutrino sector  than  predicted by the Standard Model, or alternatively, by 
 the existence of 
 an additional, light particle specie. 
 Such particles 
  may in principle be produced via a variety of mechanisms, including thermal freeze-out and non-thermal production from the decay of some heavy particle. 
Thermally produced light, weakly interacting particles  that decouples at time $t=t_{\rm d.r.}$ before neutrino decoupling $t_{\rm d.r.}< t_{\nu}$, contribute to the dark radiation density by, 
\be
\dn = c \left( \frac{g_{\star}(t_{\nu})}{g_{\star}(t_{\rm d.r.})}\right)^{4/3} \, ,
 \ee
 where $c=1,2, 4/7$ for Majorana fermion,  Dirac fermions, and scalars, respectively. Here $g_{\star}(t)$ denotes the effective number of thermal relativistic degrees of freedom  at time $t$, e.g.~at the time of neutrino decoupling, $g_{\star}(t_{\nu})=10.75$. For example, sterile neutrinos that decouple simultaneously with the Standard Model neutrinos contribute with $\dn=1$ (by construction), while 
 a very weakly interacting light particle  that decouples before $t\bar t$-annihilation when $g_{\star}(t_{\rm d.r.}) \geq 106.75$, contributes  with,
 \be
 \dn \leq 0.047 c \, .
 \ee
It follows that a  detection  of $\dn>0$ would  not directly constrain the spin of the dark radiation particle, even in the relatively simple context of thermal dark radiation. 
For a review of the predictions of various thermal models, see e.g.~\cite{brust_new_2013, steigman_equivalent_2013}.

For non-thermally produced dark radiation, there is no general relation between $\dn$ and spin, but we note that for dark radiation arising from the decay of a  field
$\Phi$ of mass $m_{\Phi}$ which reheats the universe and which may also decay into some light hidden sector scalars $\phi$ and fermions $\psi$ through the Lagrangian,
\be
\frac{\Phi}{\Lambda} \left(\frac{1}{2} \partial_{\mu}\phi \partial^{\mu}\phi + c_1 i \bar \psi \slashed{\partial} \psi + c_2 m_{\psi} \bar \psi \psi \right) \, , \label{eq:decay}
\ee 
the corresponding  decay rates are given by, 
\be
\Gamma_{\Phi \rightarrow \bar \psi \psi} = \frac{(c_1+ c_2)^2}{8\pi} \frac{m_{\psi}^2 m_{\Phi}}{\Lambda^2} \, , 
\ee
for $m_{\psi}\ll m_{\Phi}$, and 
\be
\Gamma_{\Phi \rightarrow  \phi \phi } = \frac{1}{64 \pi} \frac{m^3_{\Phi}}{ \Lambda^2} \, ,\label{eq:scalarDecayRate}
\ee
for $m_{\phi} \ll m_{\Phi}$. Thus, the decay channel into light hidden sector scalars tend  to dominate over decays into hidden sector fermions. This generic expectation has been found to be realised in several stabilised string models  
in which moduli decay into the visible sector and light hidden sector 
generically produce pseudo-scalar dark radiation \cite{cicoli_dark_2012, higaki_dark_2012, hebecker_dark_2014, angus_dark_2014}.  The amount of dark radiation produced at reheating is then given by,
\be
\dn = \frac{43}{7} \left( \frac{g_{\star}(T_{\nu})}{g_{\star}(T_{rh})}\right)^{1/3} \frac{\Gamma_{\Phi \rightarrow {\rm d.r.}}}{\Gamma_{\Phi \rightarrow {\rm vis.}}} \, .
\ee  
Dark radiation directly produced from moduli decay never thermalise and  have a characteristic energy $E_{\phi} \sim \left(\frac{M_{Pl}}{m_{\Phi}} \right)^{1/2} T_{CMB} \gg T_{CMB}$, which should be contrasted with the  general prediction of thermal models of dark radiation,  $E_{\phi} \approx T_{\phi} \lesssim T_{\rm CMB}$.

Thus,   interesting values for $\dn$ can be obtained either  through thermal production or non-thermally, and to observationally discriminate between different models, additional observables need to be considered. 
Potentially, the close proximity of the CMB spectrum to a perfect blackbody  provide 
one such observable, and in this paper we  derive the corresponding general prediction for spin-0 dark radiation.

As we will review in \S\ref{sec:mu}, heat transfer into the thermal  plasma before
 recombination
can
induce deviations of the CMB spectrum from the Planck spectrum. In particular, if the plasma is heated after the time when photon number changing processes cease to be efficient  but while electrons and photons are still tightly coupled by Compton scattering, then the   injected energy 
 distorts the CMB spectrum into a Bose-Einstein distribution with a chemical potential, $\mu$. Later energy injections into the plasma result in Compton $y$-type spectral distortions. No such distortions have been observed, 
 and the FIRAS experiment on the COBE satellite constrained the $\mu$-distortion parameter to be,  
\be
|\mu|< 9 \cdot 10^{-5} \label{eq:muBound} \, ,
\ee
 and $y< 1.5\cdot10^{-5}$ \cite{wright_interpretation_1994, fixsen_cosmic_1996}. By combining COBE/FIRAS data with data from the TRIS experiment, a somewhat tighter bound, $|\mu| < 6 \cdot 10^{-5}$, was derived in \cite{zannoni_tris_2008, gervasi_tris_2008}.
Next generation experiments may lower these limits by  3--4 orders of magnitude \cite{kogut_primordial_2011}.

To illustrate how bounds on spectral distortions may constrain a general sector of dark radiation, we will for the purpose of this introduction (and this introduction only) consider a simplified model in which dark radiation deposits energy into the plasma by scattering, 
\be
 Q = \rho_{\rm DR} \Gamma_{\rm DR \rightarrow plasma}  \, .
\ee
The resulting CMB spectral distortion parameter $\mu$ is well approximated by
\cite{sunyaev_r._small_1970},
\be
\mu = 1.4 \int_{0}^{t_f}dt \frac{Q(t)}{\rho_{\gamma}(t)} {\cal J}(t) \, , \label{eq:mueqn}
\ee
where ${\cal J}(t)$ denotes a `distortion visibility function' which we will in this introduction take to be equal to a unit-height step-function between $t_i = t(z=2\cdot 10^6)$ and $t_f = t(z=5\cdot 10^4)$. The $\mu$-parmater induced by the scattering of dark radiation is then given by, 
\be
\mu \approx 1.4 \frac{\rho_{DR}(t_{th})}{\rho_{\gamma}(t_{th}) }\int_{t_i}^{t_f}dt~\Gamma_{\rm DR \rightarrow plasma}  
= 9.2 \cdot 10^{31} \Delta N_{\rm eff}  \frac{\langle \Gamma_{\rm DR \rightarrow plasma} \rangle_t}{ {\rm GeV}} \, .
\ee
More illuminating, the COBE/FIRAS+TRIS bound $|\mu| < \mu_{FIRAS} = 6\cdot 10^{-5}$ can  be expressed as a bound on the scattering cross-section as,
\be
\dn \frac{\langle \Gamma_{\rm DR \rightarrow plasma}\rangle_t}{H_{i}} <
1.2\cdot10^{-5}~~{\rm [COBE/FIRAS]} \, .
\ee
where we have introduced  $H_i = H(z=2\cdot 10^6) = 5.5\cdot 10^{-23}$~eV.  
Thus, bounds on spectral distortions can give strong direct constraints on scattering rates and couplings between dark radiation and matter.

The purpose of this paper is to show that such constraints are harmless if the dark radiation sector consists of either scalars of pseudo-scalars.  To do so, we write down the effective theory for a spin-0 particle 
coupled to the thermal plasma with interaction strengths subjected to current, independent observational constraints. 
We then compute the dominant scattering rates of inverse Primakoff scattering, elastic scalar-Compton scattering and pair production. Using these rates, we evaluate  the heat exchange between the plasma and the dark radiation and determine  the corresponding spectral $\mu$-distortion via equation \eqref{eq:mueqn}. 
Our results are simple to state: for any physically motivated model of spin-0 dark radiation, 
\be
\mu_{\rm max}<10^{-11} \, .
\ee
Late-time spectral distortions are similarly  undetectably small. Thus, spin-0 dark radiation  produces 
 spectral distortions that are small enough to escape detection even by the next generation experiments, and a possible future detection of CMB distortions by such experiments must therefore be explained by some different physics.

Before concluding this introduction, we would like to comment on how this paper relates to previous works on spectral distortions from  axion-like particles (ALPs). It has long been appreciated that axion-like particles can convert into photons in the presence of a classical background magnetic field \cite{sikivie_experimental_1983, sikivie_detection_1985}. If one  assumes the existence of a primordial magnetic field on cosmological scales, photons may convert into ALPs on their way from the surface of last scattering to us, thereby causing  distortions of the CMB \cite{yanagida_resonant_1988, mirizzi_photon-axion_2005, mirizzi_constraining_2009,  ejlli_CMB_2013, tashiro_constraints_2013, Queiroz:2014ara}. Decaying massive axions may also contribute to the spectral distortions, as discussed in \cite{cadamuro_cosmological_2011, ejlli_bounds_2014}. Furthermore, in the presence of primordial magnetic fields and ALP dark radiation, additional constraints can be derived \cite{higaki_cosmological_2013}. 
In all cases, the ALP-photon conversion probability is very sensitive to the magnitude of the primordial cosmic magnetic field (with $P \sim B^2$ in the simplest case), which is unknown and observationally only bounded from above, c.f.~e.g.~\cite{planck_collaboration_planck_2013}. Thus, if an ALP  is discovered by other means, CMB spectral distortions may be used to better constrain the primordial magnetic field. 

In contrast, while ALP-plasma scattering is subdominant in the presence of a large classical magnetic field, the process remain active even in the absence of such a field, and  in this paper we compute the relevant cross-sections and scattering rates. Our results stresses the `darkness' of ALP dark radiation: relativistic, weakly coupled pseudo-scalars may exist in abundance in our universe, yet leave few detectable traces for experimental and observational searches to target.  These results are consistent with those obtained from a study of the effect of ALP dark radiation on BBN  \cite{conlon_cosmophenomenology_2013}, and further highlights the special status of potential signals of ALP dark radiation, such as ALP-photon conversion in galaxy cluster magnetic fields \cite{conlon_searching_2013, angus_soft_2013}. 
Conversely, our results imply that a hypothetical future detection of CMB spectral distortions (in the absence of evidence of large cosmic magnetic fields) would 
call for additional physics beyond that  of spin-0 dark radiation.

This paper is organised as follows: in \S\ref{sec:mu}, we review how energy depositions into the thermal plasma may result in spectral distortions of the CMB, in \S\ref{sec:darkrad} we write down the effective Lagrangian for baryons, electrons and photons coupled to a light scalar or pseudo-scalar particle. In \S\ref{sec:comp} and \S\ref{sec:comp2}, we compute the dominant scattering rates for the inverse Primakoff, Compton, and pair production processes. In \S\ref{sec:results}, we compute the corresponding distortions of the CMB and discuss our results.

\section{\label{sec:mu} Thermalisation and CMB spectral distortions}
\label{sec:review} 
The close proximity of the CMB spectrum to a perfect black body indicates that the thermal plasma in the early universe was close to thermal equilibrium well before recombination at $z_{\rm rec} \approx 1100$. In this section, we briefly review how an imperfect thermalisation  can lead to 
distortions of the CMB. There are many excellent and more detailed accounts of these processes, and our brief review follow  \cite{hu_thermalization_1993, partridge_3_2006}.
Recently, the physics of CMB spectral distortions has attracted much attention as the possibility of a new generation of experiments which may significantly improve the bounds by COBE/FIRAS have been considered \cite{kogut_primordial_2011}, see e.g.~\cite{Chluba_2011hw, pajer_hydrodynamical_2012, Chluba_2012gq, chluba_refined_2013, khatri_forecasts_2013, sunyaev_unavoidable_2013, amin_probing_2014,  tashiro_cmb_2014}.


After $e^{+}e^{-}$ annihilation,  the universe was  filled with a hot and  
 tightly coupled
 plasma consisting of photons, electrons and baryons. Compton scattering, $e+\gamma  \rightarrow e+ \gamma$, of photons off electrons was by far the dominant interaction and served to re-distribute photon energies. With only Compton scattering contributing to the collision operator
 of
 the Boltzmann equation, the 
 photon occupation number, $f(t, x_e )$,
 is 
%
 determined by the
 Kompaneets equation \cite{kompaneets_a.s.__1957, kompaneets_a.s.__1957-1},
 \be
 \left( \frac{\partial f}{\partial t}\right)_K = \frac{1}{t_K} \frac{1}{x_e^2} \frac{\partial }{\partial x_e} \left[ x_e^4\left( \frac{\partial f}{\partial x_e} + f + f^2 \right) \right]\, ,  \label{eq:Komp}
 \ee
 where $x_e = \omega/T_e$ for the electron temperature $T_e$, and where, 
 \be
 t_K = 9.81 \cdot 10^{27} \left(1- \frac{Y_p}{2}\right)^{-1} (\Omega_B h^2)^{-1} \left(\frac{T}{T_e} \right) z^{-4}~{\rm s} \, .
 \ee
From equation \eqref{eq:Komp} it follows that any induced deviation from a thermal spectrum evolves into to a   Bose-Einstein distribution of the form,
 \be
f_{\mu}(\omega) = \frac{1}{e^{x_e + \mu} -1} \, , \label{eq:mu}
\ee  
with, in general,  a non-vanishing and frequency dependent chemical potential $\mu(x)$. The relaxation of the photon spectrum into the thermalised Planck spectrum with $\mu =0$ then requires that additional, photon changing processes act efficiently over the expansion time-scale. 

The leading photon-number changing processes in the plasma are
(inelastic) double-Compton scattering, $e+\gamma \rightarrow e+ \gamma +\gamma$, and (less importantly) 
 bremsstrahlung,  $e+X \rightarrow e+X+\gamma$. Taking these processes into account, the full evolution of the photon  occupation number is given by,
\be
 \left( \frac{\partial f}{\partial t}\right) =  \left( \frac{\partial f}{\partial t}\right)_K +  \left( \frac{\partial f}{\partial t}\right)_{\rm br} +  \left( \frac{\partial f}{\partial t}\right)_{\rm DC} \, ,
\ee
as in reference \cite{hu_thermalization_1993}. The photon-number changing processes allow for an evolution of the $\mu$-parameter  according to the  differential equation \cite{sunyaev_r._small_1970}, 
\be
\frac{d\mu}{dt} = -\frac{1.7}{\phi(\mu) M(\mu)} \int_0^{\infty} dx_e x_e^2 \left[ \left(\frac{\partial \eta}{\partial t}\right)_{\rm br} + \left(\frac{\partial \eta}{\partial t}\right)_{\rm DC} \right] \, , \label{eq:muevol}
\ee
where we have followed the notation of \cite{burigana_c.__1991, partridge_3_2006} and denoted
\bea
\phi(\mu) &=& \frac{1}{2.404} \int_0^{\infty} dx_e x_e^2 \left(\frac{1}{e^{x_e + \mu} -1}\right) \, , \\
M(\mu) &=&\frac{d }{d \mu} \Big[ 3 \ln f(\mu) - 4 \ln \phi(\mu) \Big] \, , \\
f(\mu) &=& \frac{1}{6.494} \int_0^{\infty} dx_e x_e^3 \left( \frac{1}{e^{x_e+\mu} -1}\right) \, .
\eea
Here $\left(\frac{\partial \eta}{\partial t}\right)_{\rm br}, \left(\frac{\partial \eta}{\partial t}\right)_{\rm DC}$ denotes the production rate of new photons from bremsstrahlung and double Compton scattering, e.g.
\be
\left(\frac{\partial \eta}{\partial t}\right)_{\rm DC} \approx 3.3\cdot 10^{-39} \left(\frac{T}{T_e} \right)^{-2} f(\mu) \left( \Omega_b h^2\right) (1+z)^5 \frac{1- \eta(e^{x_e} -1)}{x_e^3 e^{3x_e/2}} \, .
\ee
Numerical studies of
\eqref{eq:muevol} have shown that arbitrarily large distortions of the photon spectrum relax back to the full thermalised spectrum as long as these distortions are introduced sufficiently early, more specifically at red-shifts $z>z_{\rm th}$  where \cite{danese_l.__1982, partridge_3_2006},
\be
z_{\rm th} = 1.8\cdot 10^6 \left( \Omega_b h^2 \right)^{-0.36} \, .
\ee
Conversely, for perturbations to the photon spectrum induced at redshifts $z<z_{th}$, the photon number changing processes cannot efficiently relax the chemical potential to zero, and such perturbations may leave an imprint on the CMB in the form a spectrum with a non-vanishing $\mu$-parameter. 

At low enough red-shifts,
\be
z< z_{\rm C} \approx 2.15 \cdot 10^4 \left(\Omega_B h^2\right)^{-1/2} \, ,
\ee
Compton scattering ceases to be efficient in establishing kinetic equilibrium. 
Heat deposited into the electrons at $z_{\rm rec} < z< z_{\rm C}$ 
%
source  Compton $y$-type  spectral distortions of the CMB. Intermediate type distortions which are not well-described by superimposed $\mu$-type and a $y$-type distortion have been considered in \cite{Hu_1995em, Chluba_2011hw, khatri_beyond_2012}. 
In \S\ref{sec:results}, we will see that the dominant contributions to spectral distortions from scattering arise from processes which are not kinematically accessible at low red-shifts, and we will therefore 
in this paper focus on $\mu$-type spectral distortions.



Intrinsic photon production is inefficient in the plasma during the $\mu$-era, and  spectral distortions are sourced by any deposited  heat which is not accompanied by the appropriate change in photon number.  More precisely, a sufficient criterium\footnote{See \cite{Chluba_2014wda} for a discussion on necessary criteria for the generation of spectral distortions. } for the generation of spectral distortions is  from some deposited heat, $\Delta \rho_{\gamma}$, and photons, $\Delta n_{\gamma}$, is,  
\be
\frac{\Delta \rho_{\gamma}}{\rho_{\gamma}} \neq \frac{4}{3} \frac{\Delta n_{\gamma}}{n_{\gamma}} \, .
\ee 
Under the assumption that double-Compton scattering is the dominant photon number changing interaction (which is well-motivated at high enough frequency), a simplified formulae for the present-day $\mu$-parameter sourced by some injected heat $Q(t)$ 
 was derived in 
\cite{sunyaev_r._small_1970, hu_thermalization_1993},
and is given by,
\be
\mu = 1.4 \int_{0}^{t(z_C)}dt \left( \frac{Q(t)}{\rho_{\gamma}(t)} - \frac{4}{3} \frac{\dot n_{\gamma}}{n_{\gamma}} \right) {\cal J}(t) \, , \label{eq:musol}
\ee
where the `distortion visibility function' ${\cal J}$ is to a good approximation given by ${\cal J}(z) = e^{-(z/z_{\mu})^{5/2}}$ \cite{chluba_refined_2013} with,
\be
z_{\mu} =2.0\cdot 10^6 \left( \frac{\Omega_b h^2}{0.02}\right)^{-2/5} \left(1- \frac{Y_p}{2}\right)^{-2/5}  \, .
\ee
In \S\ref{sec:results} we will solve \eqref{eq:musol} to obtain the $\mu$-parameter for heat and photon number changing processes sourced by scalar dark radiation. 

One additional aspect of the thermalisation process will be important for us: 
at very low frequencies, $x< x_{\rm crit.}$,  the double Compton and bremsstrahlung scattering rates become 
competitive with elastic Compton scattering, and the spectrum is quickly returned to the Planck distribution. The frequency at which the rates become equal is red-shift dependent and given by,
\be
x_{\rm crit.}^2(z) = x_{\rm br.}^2 + x_{\rm dC}^2 \, ,
\ee
where,
\bea
x_{\rm br. } &\approx& 12(1+z)^{-3/4}  \, ,\\
x_{\rm dC} &\approx& 3\cdot 10^{-6}(1+z)^{1/2}  \, .
\eea
 In \S\ref{sec:results}, we will account for this effect by restricting the range of red-shifts for each mode, $\omega(z)$, to those which satisfy $\omega(z) > x_{\rm crit.}(z) T_{CMB}(z)$.


\section{\label{sec:darkrad} Scattering of spin-0 dark radiation}
In this section, we construct the effective theory for scalar and pseudo-scalar dark radiation coupled to the thermal plasma. We then proceed to compute the relevant scattering rates for pair production and absorption processes. 

\subsection{Effective field theory of dark radiation}
We consider a low-energy effective theory (EFT) consisting of
a hypothetical sector of spin-0 dark radiation, $\phi$, and the tightly coupled thermal plasma 
of H and $^4$He baryons, electrons, and photons.   The EFT description is applicable for processes with $\sqrt{s} < \La{EFT} \approx \La{QCD}$, beyond which QCD effects have to explicitly be taken into account. The full Lagrangian may be written as,
%
\be
{\cal L}_{\rm tot} = {\cal L}_{\rm vis.}+{\cal L}_{\rm d.r.} + {\cal L}_{\rm int.} \, ,
\ee
were the  renormalisable contribution to the visible sector Lagrangian is, 
\be
{\cal L}_{\rm vis.} = - \frac{1}{4} F_{\mu \nu} F^{\mu \nu} +  \sum_{i} \bar \psi_i \left( i \slashed D + m_i\right) \psi_i \, . \label{eq:Lvis} 
\ee
Here the index $i$ in principle runs over all the visible sector fermions, which at energies below $\La{EFT}$ include the leptons 
 and the baryons. 
In \S\ref{sec:comp} and \S\ref{sec:comp2} we will consider scattering of $\phi$ off particles which are already present in  the plasma as well as particles $f$ which are kinematically accessible through the pair production process $\phi+\gamma \rightarrow f+\bar  f$. 
Thus, we take the sum over $i$ to run over electrons, muons,  and the H and $^4$He baryons.
%
We note that muon pair production will only be relevant for processes $\sqrt{s}$ close to the cut-off of the EFT, beyond which the hadronisation processes following quark/anti-quark production  to meson final states will significantly modify the analysis.\footnote{It does not appear implausible to us that for processes accessing centre-of-mass energies beyond \La{EFT},
 hadronisation result in a partial thermalisation of the secondaries and small spectral distortions as a result.} 

In this EFT, we omit the Standard Model neutrinos and the mesons. The neutrinos  are decoupled from the plasma during the period of interest and will be unimportant for our discussion of spectral distortions, and the   mesons will only contribute through pair production processes at $\sqrt{s} \gtrsim \La{EFT}$, and can safely be neglected.

The scalar sector of dark radiation  has a Lagrangian,
\be
{\cal L}_{\rm d.r.} = \frac{1}{2} \partial_{\mu} \phi \partial^{\mu} \phi - 
V_{\rm d.r.}(\phi) \, .
\label{eq:Ldr}
\ee
 We will  make the well-motivated assumption that $\phi$ has very weak self-interactions, and 
  --- since we are studying dark radiation before recombination --- 
  we will only consider the highly relativistic limit in which 
 $m_{\phi}/E_{\phi}(z_C) \ll1$. In practice, this means that we neglect $V_{\rm d.r.}(\phi)$ and only consider massless scalars. 

Famously,  light spin-less bosons tend to be technically unnatural unless protected by an approximate  symmetry which becomes exact as the mass is taken to zero.
Supersymmetry is one example of such a symmetry, but the lack of superpartners with masses close to those of the Standard Model particles indicate that supersymmetry --- if  at all realised in nature --- must be broken at a comparatively high scale and cannot protect the potential of a light scalar. 
Alternatively, 
 $\phi$ is technically natural if it appears as a (pseudo-)Nambu-Goldstone boson arising of a spontaneously broken  (approximate) global shift symmetry, $\phi \rightarrow \phi+{\rm const.}$. For such 
axion-like particles (ALP's), the shift-symmetry ensures that the field only has derivative couplings and that 
quantum effects  only generate contributions to the mass proportional to the breaking of the symmetry, which may naturally be small.  
The leading order non-renormalizable operators mediating interactions between an ALP and the particles in the thermal plasma is given by,
\be
{\cal L}_{\rm int.} = -\frac{\phi}{4 \Lambda_1}  F_{\mu \nu}  \tilde F^{\mu \nu} - \sum_{i} \frac{m_i}{\Lambda_{2 i}} \phi \bar  \psi_i  \gamma_5\psi_i \, , \label{eq:Lintps}
\ee
where we have written the interaction term $\frac{\partial_{\mu}\phi}{2\Lambda} \bar \psi \gamma^{\mu} \gamma_5 \psi$ in its non-derivative form.\footnote{See \cite{raffelt_bounds_1988, carena_effective_1989, choi_invisible-axion_1989} however, for a discussion of a  subtle effect involving multiple Nambu-Goldstone bosons in which these expression are not equivalent. 
This subtlety will not affect our subsequent discussion. }

The suppression scales $\Lambda_1= g_{a\gamma}^{-1}$ and $\Lambda_{2 i} = g_{ai}^{-1}$ are subject to stringent  astro-physical and laboratory constraints (see e.g.~\cite{beringer_review_2012} for a review). Bounds obtained from globular cluster stars require the axion-photon suppression scale to be \cite{ayala_improved_2014}\,
\be
  \Lambda_1 > 1.7\cdot 10^{10}~{\rm GeV} \, ,\label{eq:L1}
  \ee
and the coupling to electrons to be suppressed by
\be
\Lambda_{2 e} > 1.0\cdot 10^{9}~{\rm GeV} \, .  \label{eq:L2e}
\ee
Restrictions on the ALP-Compton scattering off $^4$He bounds the coupling to nucleons, 
\be
\Lambda_{2 n} > 2.3 \cdot 10^{10}~{\rm GeV} \, . \label{eq:L2n}
\ee
Finally, we note that the suppression strength for ALP-muon interactions is in comparison less constrained \cite{brust_new_2013},
\be
\Lambda_{2\mu} \gtrsim 2 \cdot 10^6~{\rm GeV} \, .   \label{eq:L2m}
\ee 
In this paper we do not consider the lepton number changing operator, $\frac{\partial_{\mu} \phi}{\Lambda_{2\mu e}} \bar \mu \gamma^5 \gamma^{\mu} e$, which is more constrained than the coupling to either type lepton,  $\Lambda_{2\mu e} \gtrsim 1.6 \cdot 10^9$~GeV \cite{beringer_review_2012}.

For the sake of completeness, we will also consider the couplings of unprotected scalars which may interact with the particles in the plasma through the `dilaton-like' and Yukawa-type  operators, 
\be
{\cal L}_{\rm int.} = -\frac{\phi}{4 \Lambda_3}  F_{\mu \nu}  F^{\mu \nu} - \sum_{i} \frac{m_i}{\Lambda_{4 i}} \phi \bar \psi_i \psi_i \, . \label{eq:Lints}
\ee
We emphasise that, in addition to being technically unnatural, such light scalar fields may cause severe cosmological problems which we do not address in this paper. For 
scalars, the Globular cluster constraints require \cite{beringer_review_2012},
\be
\Lambda_3 > 1.6 \cdot 10^{10}~{\rm GeV} \, .  \label{eq:L3}
\ee
Stronger constraints on $\Lambda_3$ can be obtained by considering the cosmological evolution of the vacuum expectation value of $\phi$. In particular, the non-observation of variations in the fine-structure constant  impose stringent bounds (for a review see \cite{damour_gravitation_????}). Furthermore, the effect of fluctuations of the fine-structure constant on the CMB is discussed in \cite{sigurdson_spatial_2003, ade_planck_2014}. For the purpose of this paper we will simply note that even for couplings which are close to saturation of the bound \eqref{eq:L3}, the scattering rate of scalar dark radiation will be too small to significantly distort the CMB blackbody spectrum.

Finally, we note that the coupling to electrons is constrained by
\be
\Lambda_{4 e} > 3.9 \cdot 10^{10}~{\rm GeV} \, . \label{eq:L4e}
\ee 
 and the coupling to nucleons is bounded by,
\be
\Lambda_{4 n} > 2.3 \cdot 10^{10}~{\rm GeV} \, . \label{eq:L4n}
\ee
 In addition, analogously to equation \eqref{eq:L2m} we take, 
 \be
 \Lambda_{4 \mu} \gtrsim 2 \cdot 10^6~{\rm GeV} \, , \label{eq:L4m}
 \ee
as an approximate constraint on the scalar-muon coupling.

In section \S\ref{sec:results} we will show that even with the strongest, non-excluded couplings to matter,   any physically motivated model of ALP or scalar dark radiation  does not give rise to a significant distortion of the CMB spectrum.

\subsection{\label{sec:scattering1} \label{sec:comp} Pseudo-scalar-plasma scattering}
In this section we compute the cross-sections of the leading interactions of an ALP with  the plasma. The relevant processes are pair production (c.f.~figure \ref{fig:pairprod}), Compton-like scattering (c.f.~figure \ref{fig:Compt}) and Primakoff scattering (c.f.~figure \ref{fig:Prim}). 

For the sake of clarity, we will here state the notation conventions used throughout this section and \S\ref{sec:comp2}. 
A scalar or pseudo-scalar field is denoted by $\phi$ and is considered to be massless and weakly coupled. 
For the purpose of this section, 
 the energy distribution of $\phi$ is arbitrary, but we note that in \S\ref{sec:results} we will only consider a mono-energetic population of dark radiation with particle energy $E_{\phi}$. The four-momentum of an incoming or out-going scalar particle is denoted by $k_1$. Furthermore, an incoming or out-going photon, $\gamma$, has four-momentum $k_2$ so that $k_1^2 = k_2^2 =0$.  We denote a general fermion of mass $m$ by $f$ (and an anti-fermion by $\bar f$), and denote the fermion momenta by $p_1$ and $p_2$. Then, $p_1^2 = p_2^2 = m^2$. The couplings of $\phi$ to electromagnetism are suppressed by the scales $\Lambda_1$ and $\Lambda_3$ as in equations \eqref{eq:Lintps} and \eqref{eq:Lints}. The couplings of $\phi$ to matter is suppressed by either $m/\Lambda_{2f}$ or $m/\Lambda_{4f}$.  While our computations are performed in flat Minkowski space,  
plasma effects are considered as they become important. 

\subsubsection{Pair production}
\label{sec:pp1}
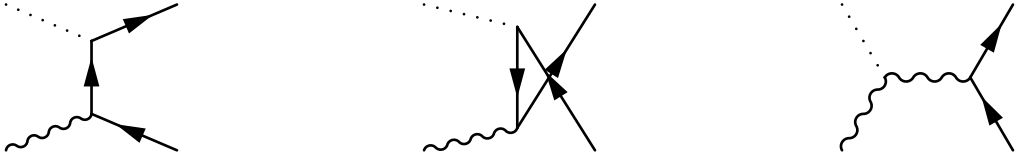
\begin{figure}[]
	\subfloat{\begin{fmffile}{cosm1}
\begin{fmfgraph*}(80,55)
\fmfleft{i1,i2}
\fmfright{o1,o2}
\fmflabel{$\gamma$}{i1}
\fmflabel{$\phi$}{i2}
\fmflabel{$~\bar f$}{o1}
\fmflabel{$f$}{o2}
\fmf{photon}{i1,v1}
\fmf{dots}{i2,v2}
\fmf{fermion}{o1,v1,v2,o2}
\end{fmfgraph*}
\end{fmffile}
\vspace*{0.5 cm}
}
	\subfloat{\hspace*{2 cm} \begin{fmffile}{cosm2}
\begin{fmfgraph*}(80,55)
\fmfleft{i1,i2}
\fmfright{o1,o2}
\fmflabel{$\gamma$}{i1}
\fmflabel{$\phi$}{i2}
\fmflabel{$\bar f$}{o1}
\fmflabel{$f$}{o2}
\fmf{photon}{i1,v1}
\fmf{dots}{i2,v2}
\fmf{phantom}{v2,o2}
\fmf{phantom}{v1,o1}
\fmf{fermion,tension=0.2}{o1,v2,v1,o2}
\end{fmfgraph*}
\end{fmffile}
}
	\subfloat{\hspace*{2 cm}
	\begin{fmffile}{Prim2}
\begin{fmfgraph*}(80,55)
\fmfleft{i1,i2}
\fmfright{o1,o2}
\fmflabel{$\gamma$}{i1}
\fmflabel{$\phi$}{i2}
\fmflabel{$\bar f$}{o1}
\fmflabel{$f$}{o2}
\fmf{photon}{i1,v1}
\fmf{dots}{i2,v1}
\fmf{photon}{v1,v2}
\fmf{fermion}{o1,v2,o2}
\end{fmfgraph*}
\end{fmffile}} \\
	\caption{The $\phi+\gamma \rightarrow f+ \bar f$ pair production processes.}
\label{fig:pairprod}
\end{figure}

The pair production process $\phi+\gamma \rightarrow f+\bar f$ tends to be the dominant interaction process between $\phi$ and the plasma at high enough ALP energies. While we here keep $f$ arbitrary, we note that   only electron and (possibly) muon pair production processes are consistent with $\sqrt{s}< \La{EFT}$.  As illustrated in figure \ref{fig:pairprod},  three diagrams contribute to this process at tree-level.  Two of these diagrams involve the ALP coupling to the fermion, and the last diagram proceed through an ALP-photon interaction. Here, we will take the interference term between the first two diagrams into account,  but we will neglect the interference between these two diagrams and the third diagram. We expect this to be a good approximation for $\Lambda_1 \not\approx \Lambda_{2f}/\sqrt{\alpha}$, and else to give 
%
a scattering rate which will differ in detail (but to the order of magnitude) from our expression.

{\bf  Pair production process 1:} We sum over final state spins and average over incoming photon polarisation to obtain  the squared amplitude of the first two diagrams, which in the notation explained in the beginning of this section is given by,
\be
\frac{1}{2} \sum_{\rm pol.} |{\cal M}|^2 =  2 e^2 \left(\frac{m}{\Lambda_{2f}}\right)^2\left(\frac{p_1\cdot k_2}{p_2 \cdot k_2}  +  \frac{p_2\cdot k_2}{p_1 \cdot k_2} + 2  \right) \, .
\ee

 The photons are Planck distributed at temperature $T$, and after integrating over 
 the angle between the incoming ALP and photon momenta we obtain the scattering rate, 
\bea
 &\Gamma_{\rm p.p.1} =\langle n_{\gamma} v \sigma \rangle =&  
    \frac{\alpha}{\pi^2}     \frac{ m^6}{E_{\phi}^3 \Lambda_{2f}^2 }\int_0^{2} \frac{d \lambda}{\lambda^3} \frac{ {\cal J}_1(\lambda)}{e^{  2m^2/(\lambda E_{\phi} T)}-1} \, ,
      \label{eq:GammaCosmophen}
\eea
where $\lambda = 2 m^2/(E_{\phi} \omega_2)\in [0,2]$ and
\be
{\cal J}_1(\lambda) =  -\sqrt{4-2 \lambda }+(\lambda -2) \log \left(\sqrt{2}-\sqrt{2-\lambda }\right)+  
2  \log \Big(\sqrt{2-\lambda } +\sqrt{2}\Big)-\frac{1}{2} \lambda  \log
   (\lambda )    \, . \nonumber 
\ee

{\bf  Pair production process 2:} The third process of figure \ref{fig:pairprod} 
proceeds through an off-shell photon. 
After summing over final state spins and averaging over initial photon polarisation 
we find the squared matrix element, 
\be
\frac{1}{2} \sum_{\rm pol.} |{\cal M}|^2 = \frac{e^2}{\Lambda_1^2} \left(m^2+\frac{(p_1\cdot k_1)^2 + (p_1\cdot k_2)^2}{k_1 \cdot k_2} \right) \, .
\ee
The scattering rate is  given by,
\be
\Gamma_{\rm p.p.2} 
\label{eq:GammaTiltedPrim} 
=
\frac{\alpha}{24 \pi^2  }  \frac{ m^6}{E_{\phi}^3 \Lambda_1^2} \int_{0}^{2}  \frac{d \lambda}{\lambda^4} \frac{ {\cal J}_2(\lambda)}{e^{  2m^2/(\lambda E_{\phi} T)}-1} 
\ee
where, 
\be
{\cal J}_2(\lambda) = 
  2(4-3 \lambda) \sqrt{4-2\lambda} + 2 \lambda^2 \ln\left( \sqrt{2} + \sqrt{2-\lambda} \right) - \lambda^2 \ln \lambda
 \, .
\ee

\subsubsection{Compton and Primakoff scattering}
\begin{figure}[]
\begin{center}
	\subfloat{\begin{fmffile}{ac1}
\begin{fmfgraph*}(80,40)
\fmfleft{i1,i2}
\fmfright{o1,o2}
\fmflabel{$f$}{i1}
\fmflabel{$\phi$}{i2}
\fmflabel{$f$}{o1}
\fmflabel{$\gamma$}{o2}
\fmf{fermion}{i1,v1,v2,o1}
\fmf{dots, tension=0.2}{i2,v1}
\fmf{photon, tension=0.2}{v2,o2}
\end{fmfgraph*}
\end{fmffile}
}
	\subfloat{\hspace*{4 cm}
\begin{fmffile}{ac2}
\begin{fmfgraph*}(80,40)
\fmfleft{i1,i2}
\fmfright{o1,o2}
\fmflabel{$f$}{i1}
\fmflabel{$\phi$}{i2}
\fmflabel{$f$}{o1}
\fmflabel{$\gamma$}{o2}
\fmf{fermion}{i1,v1,v2,o1}
\fmf{phantom}{i1,v1} 
\fmf{phantom}{v2,o1} 
\fmf{photon,tension=0.2}{v1,o2}
\fmf{dots,tension=0.2}{i2,v2}
\end{fmfgraph*}
\end{fmffile}
} 
\end{center}
	\caption{Compton-like scattering.}
\label{fig:Compt}
\end{figure}
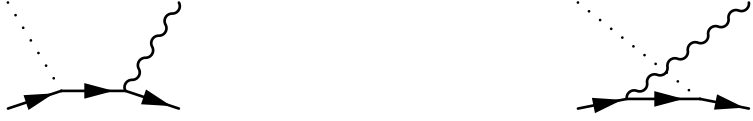

\begin{figure}
\begin{center}
\vspace*{2cm}
\begin{fmffile}{Prim}
\begin{fmfgraph*}(80,45)
\fmfleft{i1,i2}
\fmfright{o1,o2}
\fmflabel{$f$}{i1}
\fmflabel{$\phi$}{i2}
\fmflabel{$f$}{o1}
\fmflabel{$\gamma$}{o2}
\fmf{photon}{v2,o2}
\fmf{dots}{i2,v2}
\fmf{fermion}{i1,v1,o1}
\fmf{photon}{v1,v2}
\end{fmfgraph*}
\end{fmffile}
\end{center}
\vspace*{1 cm} 	
	\caption{The inverse Primakoff process.}
	\label{fig:Prim}
	\end{figure}
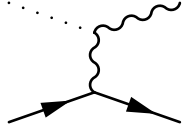

In the Compton-like process of figure \ref{fig:Compt} and the (inverse) Primakoff process of figure \ref{fig:Prim}, ALPs are absorbed in the plasma.  In this section we will compute the corresponding scattering cross-sections. In principle, the amplitude of the Compton process should be added coherently to that of the Primakoff process, but here we compute the scattering rate of the Compton process separately from that of the Primakoff process, thus neglecting an interference term which could change the detailed expression for  $\Lambda_1 \approx \Lambda_{2f}/\sqrt{\alpha}$.


{\bf Compton-like ALP-absorption:} After summing over the final state photon polarisation and averaging over the spin of the incoming fermion we find the  squared matrix element,   
\be
\frac{1}{2}\sum_{\rm pol} |{\cal M}|^2 = 2 e^2 \left(\frac{m}{\Lambda_{2f}}\right)^2 \left( \frac{p_1 \cdot k_1}{p_1 \cdot k_2} + \frac{p_1 \cdot k_2}{p_1 \cdot k_1} +2 \right) \, .
\ee
For the time-period of  interest for CMB spectral $\mu$-distortions, $T \ll m_e$, and the incoming fermion is approximately stationary in the 
rest frame of the plasma, i.e.~$p_1 \approx (m, \vec{0})$.
The 
cross-section is then given by,
\be
(\sigma v)_{\rm Compton} = \frac{\alpha }{2 \Lambda_{2f}^2} \left[ \frac{2}{1+2r} + \frac{1+r}{(1+2r)^2} + \frac{1}{r} \ln(1+ 2r) \right] \, , \label{eq:ComptGamma}
\ee
where we have introduced the notation $r= E_{\phi}/m$. 
The scattering rate from the Compton process is  given by $\Gamma_{\rm Compton} = a_f n_B(T) (\sigma v)_{\rm Compton}$ where, 
\be
n_B(T) = \eta \frac{2 \zeta(3)}{\pi^2} T^3 \, ,
\ee
denotes the baryon density, $a_f = (1- \frac{Y_p}{2}), (1-Y_p)$ and $\frac{Y_p}{4}$ denotes the fractional abundance of electrons, Hydrogen ions and Helium ions, respectively, and 
 $\eta = n_B/n_{\gamma} \approx  6.1 \cdot 10^{-10}$  \cite{nollett_bbn_2014}. Finally, we note that in the $r\rightarrow 0$ limit,
\be
(\sigma v)_{\rm Compton} = \frac{5\alpha}{2 \Lambda_{2f}^2} \left( 1+ {\cal O}(r) \right) \, .
\ee

{\bf Inverse Primakoff: } For the Primakoff processes of figure \ref{fig:Prim}, 
the amplitude-squared is given by 
\bea
\frac{1}{2} \sum_{\rm pol} |{\cal M}|^2 
&=& \frac{e^2}{\Lambda_1^2} \frac{1}{k_1\cdot k_2} \left( (k_1\cdot p_1)^2 +(k_1\cdot p_2)^2 - m^2 (k_1\cdot k_2)\right) \label{eq:primMsquare} \, .
\eea 
The $1/(k_1\cdot k_2) = 1/t$ pre-factor indicates that this cross-section has a collinear divergence in flat Minkowski space which  must be regularised. 
In higher order computations such singularities are cancelled by bremmstrahlung contributions. Here there is no such cancellation, but the cross-section is regularised by environmental effects.
In the pre-recombination thermal plasma, the   transverse polarisation of the photon obtains an effective mass through the plasma frequency,
\be
m_{\gamma} = \omega_{pl} = \sqrt{\frac{4 \pi \alpha n_e}{m_e}} \, ,
\ee
and it's tempting to conclude that this mass regulates the intermediate state on-shell divergence.  However, the correct regulator is the Debye-H\"uckel inverse radius, $k$,  outside which the plasma screens the proton or electron charge with an effective Yukawa potential \cite{raffelt_astrophysical_1986},
\be
V(r) = \frac{Ze}{4 \pi r} e^{-k r} \, ,
\ee 
where
\be
k^2 = \frac{4 \pi \alpha \tilde n}{T} \, . \label{eq:DH}
\ee 
Here $\tilde n$ denotes the weighted number of charged particles,
\be
\tilde n = n_e + \sum_{\rm \{ H,~He\}} Z_j^2 n_j = \left(2-  \frac{Y_p}{2}\right) n_B \, ,
\ee
where we in the last two steps have specialised to the primordial plasma consisting of electrons, hydrogen and helium (with a mass fraction $Y_p$).

In the co-moving laboratory frame, the regularisation of the collinear divergence results in a red-shift independent enhancement of the cross-section by,
\be
\sim \ln \left(\frac{4E_{\phi}^2}{k^2} \right)= \ln \left(\frac{4E_{\phi}^2 T}{4\pi \alpha(2 -Y_p/2) n_B}\right) = 
\ln \left(
\frac{\pi  }{ \alpha(2-Y_p/2) \eta 2 \zeta(3) }\frac{E_{\phi}^2}{T^2}
\right) \, . \label{eq:logenhancement}
\ee

For
$
E_{\phi}/T \gg 3 \cdot 10^{-6} 
$ --- which contains  the entire physically well-motivated region of parameter space ---
the log-enhanced terms dominate the cross-section. Truncating to the leading order expression, the Primakoff scattering cross-section is given by,
%
%
\bea
(\sigma v)_{\rm Primakoff} &=& 
   \frac{\alpha}{8\pi \Lambda_1^2}  \ln \left(\frac{4E_{\phi}^2}{k^2} \right) \nonumber \\
&=&
 \frac{\alpha}{8\pi \Lambda_1^2} \ln \left(
\frac{\pi E_{\phi}^2 }{ \alpha(2- Y_p/2) \eta 2 \zeta(3) T^2}
\right) \, .
\eea

\subsection{\label{sec:scattering2} \label{sec:comp2} Scalar-plasma scattering}
In this section, we will discuss the pair production and absorption processes for scalar dark radiation interacting with the plasma through the Lagrangian
\eqref{eq:Lints}. 

\subsubsection{Pair production}
{\bf Pair production process 1:} Starting with pair production through the scalar-fermion 
coupling, i.e.~considering the first two diagrams of figure \ref{fig:pairprod}, the
polarisation-summed  squared amplitude is given by,
\bea
\frac{1}{2}\sum_{\rm pol} |{\cal M}|^2 &=& 2 e^2 \frac{m^2}{\Lambda_{4f}^2} \Big[ 
\frac{p_1\cdot k_1}{p_1\cdot k_2} + \frac{p_1 \cdot k_2}{p_1 \cdot k_1} + 2 m^2\left( \frac{m^2-p_1\cdot k_1}{(p_1 \cdot k_1)^2} + \frac{m^2-p_1\cdot k_2}{(p_1 \cdot k_2)^2}\right) \nonumber \\
&+&2 +2m^2 \frac{2m^2- (p_1\cdot k_1)-( p_1\cdot k_2)}{(p_1\cdot k_1)( p_1\cdot k_2)}
\Big] \, .
\eea
The thermally averaged scattering rate in the comoving rest frame of the plasma  is then given by,
 \be
 \Gamma_{\rm p.p.1} = \langle n_{\gamma} v \sigma \rangle 
 = \frac{ \alpha m^6}{ \pi^2 \Lambda_{4f}^2 E_{\phi}^3 } \int_0^2 \frac{{\rm d}\lambda}{\lambda^3} \frac{{\cal J}_3(\lambda) }{e^{2m^2/(E_{\phi} T \lambda)}-1} \, , \label{eq:GammaCosmophens}
 \ee
where, again, $\lambda = 2m^2/(E_{\phi} \omega_2)$ and,  
\be
{\cal J}_3(\lambda)= \int_{0}^{1-\lambda}dy
 \left(\Delta(2+ \Delta - \Delta^3) + \frac{1}{2}(1- \Delta + \Delta^2 + \Delta^5) \ln \left( \frac{1+ \Delta}{1- \Delta}\right) \right)  \nonumber \, ,
\ee
with  $\Delta = \sqrt{\frac{1-y-\lambda}{1-y}}$.

{\bf Pair production process 2: } The second pair production process proceeds through the `dilation' coupling of equation \eqref{eq:Lints} and results in a squared amplitude of,
\be
\frac{1}{2} \sum_{\rm pol} |{\cal M}|^2 = \frac{e^2}{4\Lambda_3^2 k_1\cdot k_2} \left( (k_1\cdot k_2)^2 - m^2 k_1\cdot k_2 - 2 (p_1 \cdot k_1)( p_1 \cdot k_2) \right) \, .
\ee
The thermally averaged cross-section is then,
\bea
\Gamma_{\rm p.p. 2} 
&=&
 \frac{\alpha}{96 \pi^2 }  \frac{m^6}{E_{\phi}^3 \Lambda^2_3} \int_0^{2} \frac{{\rm d}\lambda}{\lambda^4}
  \frac{{\cal J}_4(\lambda)}{e^{2m^2/(\lambda E_{\phi} T)} -1}
\, , \label{eq:GammaTiltedPrims}
\eea
where,
\be
{\cal J}_4(\lambda) = 
 2(4-5\lambda) \sqrt{4-2\lambda} + 6 \lambda^2 \ln\left(\sqrt{2} + \sqrt{2-\lambda} \right)
-3 \lambda^2 \ln (\lambda)  \, .
\ee

\subsubsection{Compton and Primakoff scattering}
We now turn to the absorption processes of scalar-Compton scattering and the scalar Primakoff process. 

{\bf Compton-like scalar absorption:}  
After summing over final state polarisations and averaging over the initial spin, the 
squared matrix element for the Compton-like process is given by,
\bea
&\frac{1}{2} \sum |{\cal M}|^2& = 2 e^2 \frac{m^2}{\Lambda_{4f}^2} \left[ \frac{p_1\cdot k_1}{p_1 \cdot k_2} + \frac{p_2 \cdot k_2}{p_1 \cdot k_1} -2 +  \right. \nonumber \\
&+& \left. 2 m^2 \left( \frac{2m^2 + p_1\cdot k_1 - p_1\cdot k_2}{p_1\cdot k_1 p_1 \cdot k_2}
-
\frac{m^2 +p_1 \cdot k_1}{(p_1\cdot k_1)^2}
-
\frac{m^2 - p_1 \cdot k_2}{(p_1 \cdot k_2)^2}
\right)\right] \, .
\eea
The resulting  scattering cross-section is given by,
\be
(\sigma v)_{\rm Compton} = \frac{\alpha }{8  \Lambda^2_{4f} } \frac{1}{r^3} \left( (2+r)^2 \ln(1+2r) 
-\frac{2r(2+3r)(2+r(5+r))}{(1+2r)^2}
\right) \, . \label{eq:ComptGammas}
\ee
Here again, we have denoted $r = E_{\phi}/m$, and we  note that in the $r\rightarrow0$ limit, 
\be
(\sigma v)_{\rm Compton} = \frac{ \alpha }{3 \Lambda_{4f}^2} \left(1+{\cal O}(r) \right) \, .
\ee
{\bf Inverse scalar-Primakoff:}
The scalar Primakoff process of figure \ref{fig:Prim} has a squared matrix element of,
\be
\frac{1}{2} \sum |{\cal M}|^2 = \frac{e^2}{4\Lambda_3^2 k_1\cdot k_2} \left[ (k_1\cdot k_2)^2 - m^2 k_1 \cdot k_2 +2(p_1 \cdot k_1) (p_1 \cdot k_2)\right] \, ,
\ee
which again exhibits a collinear divergence which is regularised at the Debye-H\"uckel inverse radius. 
Considering only the  collinearly enhanced terms, the cross-section is simply given by,
\be
(\sigma v) = \frac{\alpha}{16 \Lambda_3^2}\ln( 4 E_{\phi}^2/k^2) \, ,
\ee
with  $k$ as in equation \eqref{eq:DH}. 
The scattering rate is given by,
\be
\Gamma_{\rm Primakoff} = a_f n_B(T)  (\sigma v) =  \frac{\alpha a_f n_B(T)}{16 \Lambda_3^2} \ln( 4 E_{\phi}^2/k^2) \, .
\ee

\section{\label{sec:results} Results}
We are now ready to compute the $\mu$-parameter from the scattering rates obtained in  \S\ref{sec:comp} and \S\ref{sec:comp2}.
We will throughout this section assume a mono-energetic population of dark radiation  with energy $E_{\phi}(z)$ so that the dark radiation distribution function $f_{\phi}(\omega)$  is given by,
\be
f_{\phi}(\omega) = \delta(\omega- E_{\phi}) n_{\phi} = \delta(\omega- E_{\phi}) \frac{\rho_{\phi}}{E_{\phi}} \, .
\ee

\subsection{Heat transfer and change of photon number}
\label{sec:Qs}

While pair production processes are  relevant only if the characteristic energy of the dark radiation is sufficiently high, the Compton-like process and the Primakoff process operate at all temperatures.  Furthermore, considering the diagrams of figures \ref{fig:Compt} and \ref{fig:Prim} with time running from right to left, it is immediately clear that both directions of energy exchange must be considered, i.e.~energy leakage from the plasma into the dark radiation sector may well be competitive with the dark radiation energy deposit into the plasma. 

In addition, each of the processes considered in \S\ref{sec:comp} and \S\ref{sec:comp2}  lead to a net change in photon number by $+1$ (for the pair production process we assume $\phi+\gamma \rightarrow f+\bar f \rightarrow  \gamma+\gamma$), so that  the second term of equation \eqref{eq:musol}, which we repeat here for clarity,
\be
\mu = 1.4 \int_{0}^{t(z_C)}dt {\cal J}(t)\left( \frac{ Q(t)}{\rho_{\gamma}(t)} - \frac{4}{3} \frac{\dot n_{\gamma}}{n_{\gamma}} \right) \, , \nonumber
\ee
may contribute significantly to the final value of the $\mu$-parameter.

\subsubsection{Pair production}
We will now evaluate the differential contribution to the final $\mu$-parameter from the pair production processes, $\phi + \gamma \rightarrow f +\bar f$, for scalars and pseudo-scalars. For $\sqrt{s} < \La{EFT}$, the only available final states are the electron and possibly the muon.

The rate at which heat is deposited in the plasma is given by,
\be
 Q^{\rm tot}_{\rm Pair~production}  = \Gamma^{\rm tot}_{\rm p.p.} \rho_{\phi} \, , \label{eq:energyinjection}
\ee
where $\Gamma^{\rm tot}_{\rm p.p.}(E_{\phi})$ is the sum of the scattering rates of equations \eqref{eq:GammaCosmophen} and \eqref{eq:GammaTiltedPrim} in the pseudo-scalar case, and the sum of \eqref{eq:GammaCosmophens} and \eqref{eq:GammaTiltedPrims} in the scalar case. 

The net photon number change is, 
\be
\dot n_{\gamma} = + \Gamma^{\rm tot}_{\rm p.p.} n_{\phi} \, .
\ee
Thus, the largest contribution to $\mu$ will come from the change in photon number for 
\be
\frac{E_{\phi}}{T} < \frac{4 \pi^3}{90 \zeta(3)} \approx 1.1 \, ,
\ee  
while the energy injection of equation \eqref{eq:energyinjection} dominates at higher $E_{\phi}$.

The differential change in the $\mu$-parameter
is then given by
\be
\frac{ Q(t)}{\rho_{\gamma}(t)} - \frac{4}{3} \frac{\dot n_{\gamma}}{n_{\gamma}} = \Gamma^{\rm tot}_{\rm p.p.} \frac{7}{8} \left( \frac{4}{11}\right)^{4/3} \dn \left(1- \frac{4 \pi^4}{90 \zeta(3)} \frac{T}{E_{\phi}} \right) \, .
\label{eq:MuPairProd}
\ee
To obtain the corresponding contribution to the  $\mu$-parameter, this contribution is  weighted by the distortion visibility function and numerically integrated over the entire $\mu$-epoch.

\subsubsection{Compton and Primakoff scattering}
While the pair production process considered in the previous section is unidirectional, scattering of the 
%
Primakoff and Compton-type may proceed in either direction, i.e.~we need to consider $\phi+f\leftrightarrow \gamma+f$. 
The cross-sections satisfy $\sigma v |_{\phi+f\rightarrow \gamma+f} = \sigma v |_{\gamma+f\rightarrow \phi+f}=\sigma v(\omega)$, 
so  the heat exchange from the Primakoff and Compton-like scattering channels may be written as,  
\be
Q^{\rm tot}_{\rm scattering} 
= \sum_{k} a_f n_B(T) \int {\rm d}\omega \, \omega  [\sigma v]_k(\omega) \left(f_{\phi}(\omega)  - f_{\gamma}(\omega)  \right) \, ,
\ee  
where $k$ runs over the Primakoff and Compton scattering rates. The $\phi+f\rightarrow \gamma+f$ process increases the photon number by one unit so that
\be
\dot n_{\gamma} = \sum_{k} a_f n_B(T) \int {\rm d}\omega \,   [\sigma v]_k(\omega) \left(f_{\phi}(\omega)  - f_{\gamma}(\omega)  \right) \, ,
\ee
It then follows that the differential contribution to the $\mu$-parameter is given by,
\bea
\frac{ Q(t)}{\rho_{\gamma}(t)} - \frac{4}{3} \frac{\dot n_{\gamma}}{n_{\gamma}} &=& 
 \frac{7}{8} \left( \frac{4}{11}\right)^{4/3} \dn a_f n_B(T) [\sigma v]_{\rm tot}(E_{\phi})
  \left(1+ \frac{4 \pi^4}{90 \zeta(3)} \frac{T}{E_{\phi}} \right) \nonumber \\
  &-& a_f \eta \int {\rm d}\omega\, \left(\frac{30 \zeta(3) \omega}{\pi^4 T} - \frac{4}{3} \right) [\sigma v]_{\rm tot}(\omega) f_{\gamma}(\omega) \, , \label{eq:MuScattering}
\eea
where we have denoted $(\sigma v)_{\rm tot} = (\sigma v)_{\rm Primakoff}+ (\sigma v)_{\rm Compton}$.

 \subsection{Discussion: Small spectral distortions from scalar and pseudo-scalar scattering}
 \label{sec:results:results}

\begin{figure}
\begin{center}
\includegraphics[width=0.9\textwidth]{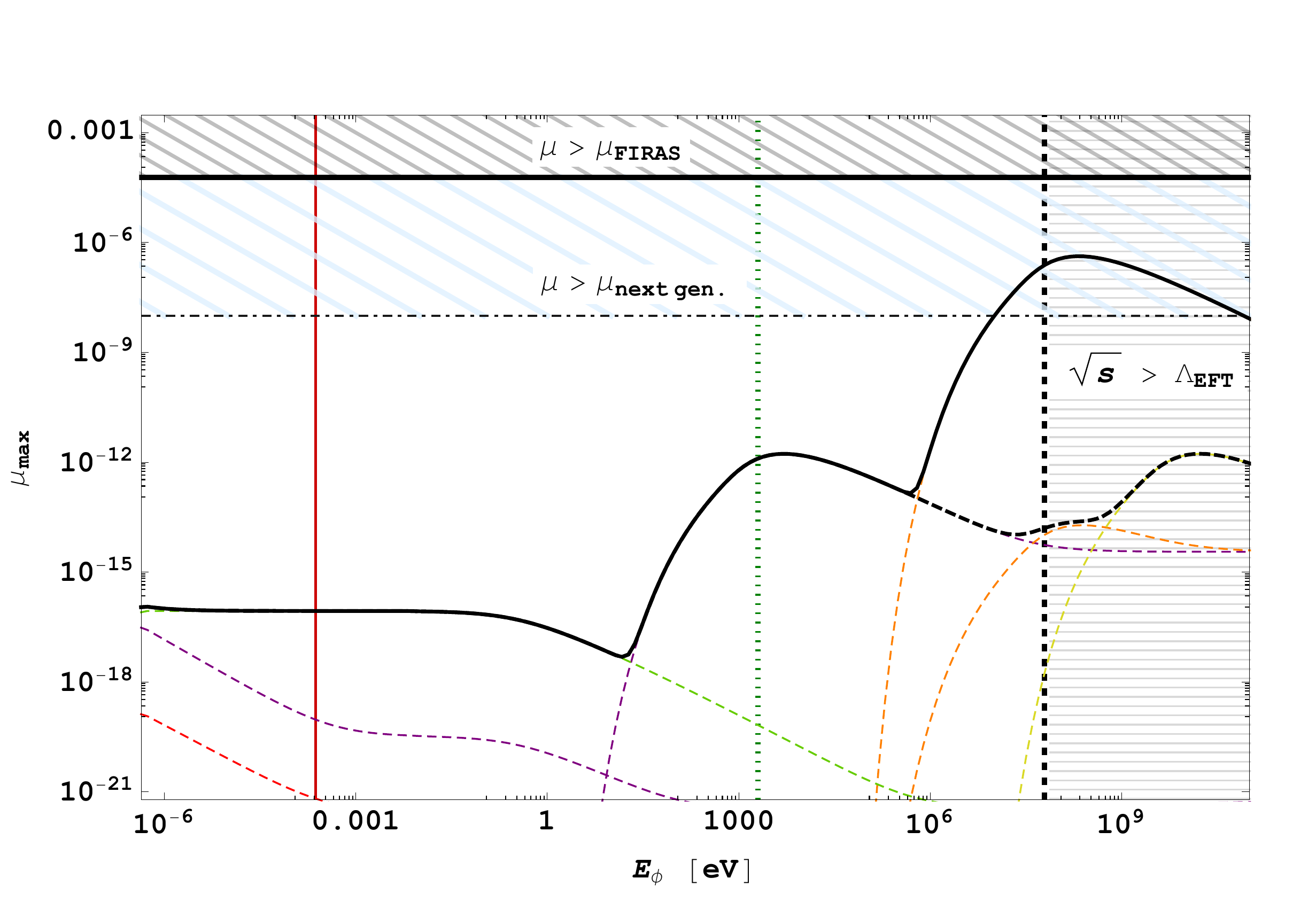}
\end{center}
\caption{Maximal obtainable values for the $\mu$-parameter for ALP dark radiation with couplings consistent with current observational bounds, c.f.~equations \eqref{eq:L1}--\eqref{eq:L2m}, and with $\dn \leq 0.79$, as given by the upper value of the  Planck  95\% confidence limit (Planck+WMAP-pol+high-l+BAO). 
The solid horizontal line indicates the COBE/FIRAS+TRIS bound, $\mu_{FIRAS} = 6 \cdot 10^{-5}$ and $\mu_{\rm next~gen.}$ indicates the potential reach of the next generation of experiments, $\mu_{\rm next~gen.} =  10^{-8}$ \cite{kogut_primordial_2011}. The solid red vertical line indicates the CMB temperature and the dashed vertical black line indicates the boundary of applicability for the EFT, $\sqrt{s} \approx \La{EFT} \approx \La{QCD}$. The dotted green vertical line indicates the maximum attainable energy for the non-thermal models discussed in the main text. At high energies, the pair production processes dominates. Electron-positron production peak at $E_{\phi}\sim$keV, as indicated by the dashed purple curve, and  $\mu^+\mu^-$ pair production   peak close to \La{EFT}. The upper dashed orange line indicates the muon contribution to the chemical potential for $\Lambda_{2\mu} = 2\cdot 10^6$~GeV as in equation \eqref{eq:L2m} and the lower purple dashed line corresponds to $\Lambda_{2\mu} = \Lambda_{2e} = 1.0 \cdot 10^9$~GeV. For illustrative purposes, the yellow dashed line indicates the naive expectation of the contribution from $p^+ \bar p^-$ pair production, though  this process is not accessible in our EFT. At lower energies, Compton and Primakoff scattering off electrons, hydrogen and helium dominate, as is indicated by the dashed purple, green and orange lines, respectively. The total contribution to the $\mu$-parameter  is given by the solid black line, and the  dashed black line indicates the total contribution for $\Lambda_{2\mu} = \Lambda_{2e}$.  
}
\label{fig:ALP}
\end{figure}

\begin{figure}
\begin{center}
\includegraphics[width=0.9\textwidth]{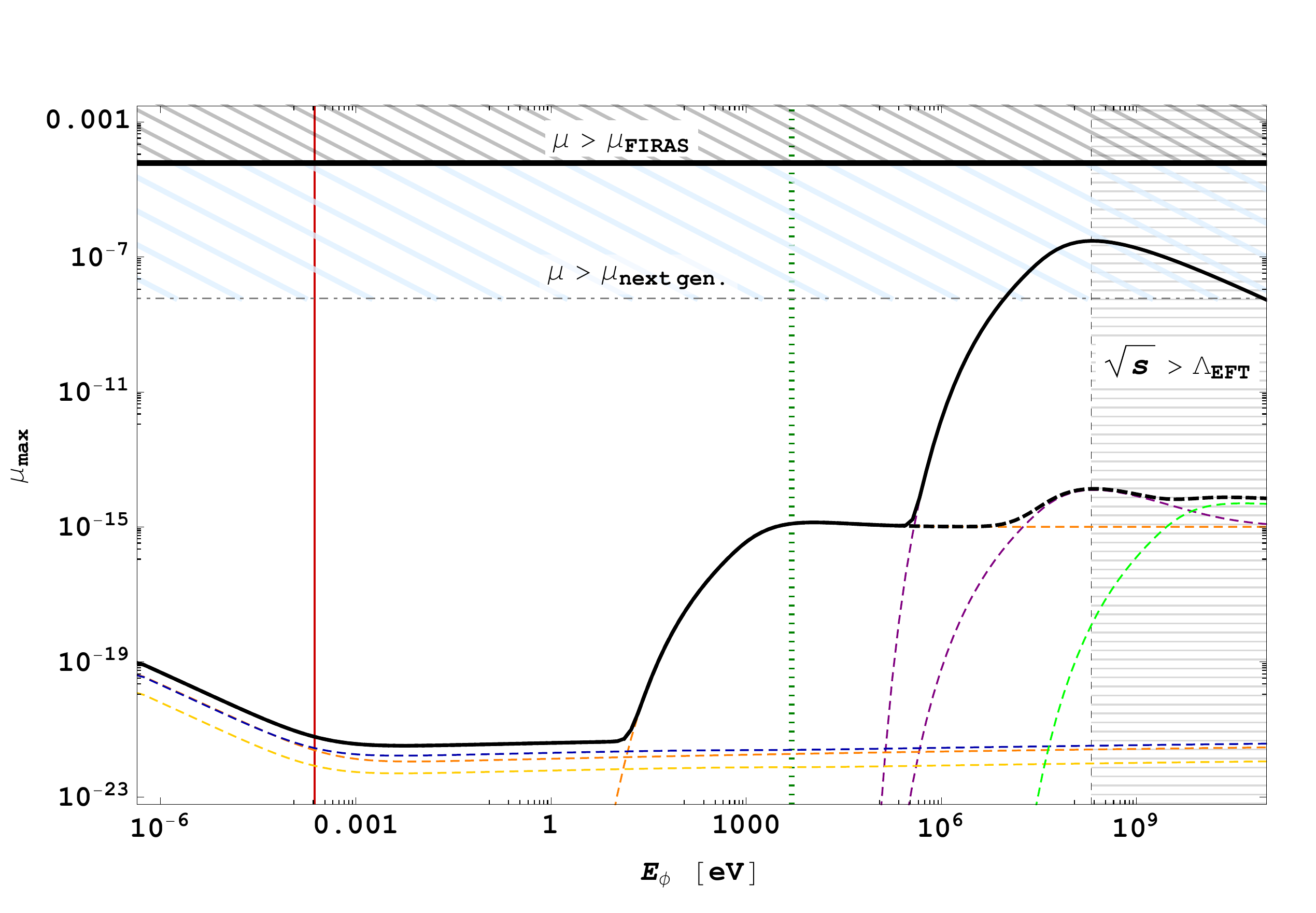}
\end{center}
\caption{Maximal obtainable values for the $\mu$-parameter for scalar dark radiation with couplings consistent with current observational bounds, c.f.~equations \eqref{eq:L3}--\eqref{eq:L4m}, and with $\dn \leq 0.79$, as given by the upper value of the  Planck  95\% confidence limit (Planck+WMAP-pol+high-l+BAO). 
The total contribution to the $\mu$-parameter is indicated by the solid black line for \La{4\mu} as in equation \eqref{eq:L4m}, and by the dashed black line for \La{4e} = \La{4\mu}.
 The remaining dashed curves are as in figure \ref{fig:ALP}.  
}
\label{fig:scalar}
\end{figure}

In this section, we evaluate the spectral distortion from the scattering processes considered in \S\ref{sec:comp} and \S\ref{sec:comp2} by numerically integrating the sum of equations \eqref{eq:MuPairProd} and \eqref{eq:MuScattering} for both scalars and pseudo-scalars. Our main results are presented in figures \ref{fig:ALP} and \ref{fig:scalar}.

Heat deposited  into the plasma at red-shift $z$ with low enough energy, $E(z)< x_{\rm crit.}(z) T_{\gamma}(z)$, does not contribute to the $\mu$-parameter, as discussed in \S\ref{sec:review}. As $x_{\rm crit}$ is $z$-dependent, this limits the range of red-shifts for which low-energy heat transfer can have an impact on $\mu$ \cite{hu_thermalization_1993}.
Hence, dark radiation with a present-day energy of $E_{\phi} < E_{\rm min} =4.3\cdot 10^{-7}$~eV does not contribute to the $\mu$-parameter at any red-shift, and provides a natural low-energy cut-off. The high-energy cut-off is provided by the EFT constraint $\sqrt{s} < \La{EFT} \approx \La{QCD}$. For the pair production processes, $\sqrt{s} \sim \sqrt{T_{\gamma}(z) E_{\phi}(z)}$.  The EFT description is then valid for $\sqrt{s} \lesssim  240$~MeV, which restricts the present-day particle energy to $E_{\phi} \lesssim E_{\rm max} = 62$~MeV. 

We now note that the  energy range,  $E_{\rm min} \leq E_{\phi} \leq E_{\rm max}$, includes all models
discussed in \S\ref{sec:intro}, in addition to a range of dark radiation effective theories that cannot be physically motivated by such models.  Thermal models of dark radiation satisfy $T_{\phi} /T_{CMB} = (g_{\star}^{\rm now}/g_{\star}(t_{\phi}))^{1/3}$. For $E_{\phi} > E_{\rm min}$ this gives,  $g_{\star}(t_{\phi}) < g_{\star}^{\rm now} (E_{\rm min}/T_{CMB})^3 = 5\cdot 10^8$, which is immediately satisfied by all generic models of thermal dark radiation. For non-thermal dark radiation produced at reheating through 2-body modulus decay (as discussed in \S\ref{sec:intro}), the present-day characteristic ALP energy is $E_{\phi} \approx (M_{Pl}/m_{\Phi})^{1/2} T_{CMB}$ \cite{conlon_cosmophenomenology_2013}, where again, $m_{\Phi}$ denotes the mass of the decaying particle. The reheating temperature is given by $T_{\rm reheat} \sim m_{\Phi}^{3/2}/M_{Pl}^{1/2}$, and by requiring that this temperature is high enough for successful BBN to proceed, i.e.~$T_{\rm reheat} >  {\cal O}(1~{\rm MeV})$, the modulus mass must satisfy $m_{\Phi} \gtrsim  30$~TeV. It then follows that $E_{\phi} \lesssim 2~{\rm keV}\ll E_{\rm max}$. We thus conclude that our effective field theory approach should be sufficient to describe the vast majority of  models of spin-0 dark radiation.  
In  figures \ref{fig:ALP} and \ref{fig:scalar} we indicate the maximal present-day particle energy for dark radiation  as produced from non-thermal particle decay by the vertical, green dashed line. While higher particle energies are captured by our effective field theory, we know of no production mechanism of such high energy dark radiation, and therefore regard them as less physically motivated. 

We will now discuss some of the prominent features of figures \ref{fig:ALP} and \ref{fig:scalar}. For low $E_{\phi}$, the pair production processes is never kinematically accessible and the Compton and Primakoff processes for scattering off  electrons, hydrogen or helium dominate.  For $E_{\phi} \ll m$, these processes have constant scattering rates (as shown in \S\ref{sec:comp}). By taking $\rho_{\rm d.r.}$  constant while lowering $E_{\phi}$, the number density of dark radiation particles increases, and so does the number of low-energy photons produced from Primakoff and Compton scattering. According to equation \eqref{eq:MuScattering}, this makes $\mu \sim 1/E_{\phi}$ at low enough energies, which explains the increase of the contribution from scattering off electrons and helium at low and decreasing energies. However, for $E_{\phi}< E_{\rm min}$, photon number changing processes  are efficient in the plasma and this low-energy increase does not enhance the spectral distortion to observable levels. 

At $E_{\phi} \approx 100$ eV, $e^-e^+$ pair production becomes kinematically accessible which increases the induced distortion somewhat, but never beyond $\mu \approx 10^{-11}$. In the well-motivated case in which $\Lambda_{2\mu} = \Lambda_{2e}$, this is also the largest induced value of $\mu$ for any dark radiation model describable in the EFT. Taking the value of  $\Lambda_{2\mu}$ to be given by equation \eqref{eq:L2m}, the spectral distortion is significantly enhanced as muon pair production becomes kinematically accessible, and may even give rise to distortions which are detectable by next generation experiments. We note however that there are three strong reasons against this case being realised: 
\begin{enumerate}
\item the constraint on $\Lambda_{2\mu}$ is comparatively weak, and is quite likely never saturated in a  realistic ALP or scalar model which simultaneously satisfies the constraints on \La{1} and \La{2e}, 
\item there is, to us, no known production mechanism of such high energy primordial dark radiation, and lacking such a mechanism, the model  is not physically well-motivated,  
\item pion production from couplings between $\phi$ and quarks may become important and may modify the distortion for $E_{\phi} \approx \La{EFT}$.   
\end{enumerate}
 
 We thus conclude that no physically motivated model of spin-0 dark radiation that would give a detectable $\mu$-distortion of the CMB through the scattering processes discussed in \S\ref{sec:comp} and \S\ref{sec:comp2}.

\section*{Acknowledgments}
I would like to thank Jens Chluba, Joseph Conlon, Pedro Ferreira, Carlos Martins, Enrico Pajer, Emanuele Re and Markus Rummel for discussions. I am particularly grateful to Jens Chluba for comments on the manuscript.  I would like to thank the organisers of the workshop `New Challenges for Early Universe Cosmologists' at the Lorentz center in Leuven, the Netherlands, for generous hospitality during the initial stage of this project.
I am funded by the European Research Council starting grant `Supersymmetry Breaking in String Theory', but the contents of this paper
reflect only the authors' views and not necessarily the views of the European Commission.


\bibliographystyle{JHEP}
\bibliography{darkness_bib}

\end{document}